\documentclass[journal]{IEEEtranTIE}
\usepackage{bm}
\usepackage{amsmath} 
\usepackage{amssymb}
\usepackage[ruled]{algorithm2e}
\usepackage{graphicx}
\usepackage{multirow}
\usepackage{pdfsync}
\usepackage{booktabs}
\usepackage[sort]{cite}

\title{Periodic Proprioceptive Stimuli Learning  and Internal Model Development for Avian-inspired  Flapping-wing Flight State Estimation}
\author{Chen Qian\(^\dagger\), Jiaxi Xing\(^\dagger\), Jifu Yan, Mingyu Luo, Shiyu Song, Xuyi Lian, \\ Yongchun Fang, Fei Gao, and Tiefeng Li\thanks{This work was supported by National Natural Science Foundation of China(T2125009),  "Pioneer" R\&D Program of Zhejiang (Grant no. 2023C03007). The authors marked with \(^\dagger\) contribute equally to this work(Corresponding authors: Yongchun Fang, Fei Gao, and Tiefeng Li).\\
Chen Qian is with Center for X-mechanics of Zhejiang University (email: chainplain@mail.nankai.edu.cn).\\ 
Jiaxi Xing is with Huzhou Institute
of Zhejiang University (email: xingjx@hizju.org).\\ 
Jifu Yan and Yongchun Fang are with Institute of Robotics and Automatic Information
Systems, College of Artificial Intelligence, Nankai University (email: yanjf2000@mail.nankai.edu.cn, fangyc@nankai.edu.cn). \\
Mingyu Luo is with School of Mechanical Engineering Zhejiang University (email: luomy@zju.edu.cn). \\
Shiyu Song, Xuyi Lian and Tiefeng Li are with School of Aeronautics and Astronautics, Zhejiang University(email: 12434023@zju.edu.cn, liam.lianxy@zju.edu.cn, litiefeng@zju.edu.cn).\\
Fei Gao is with College of Control Science and Engineering, Zhejiang University (email: fgaoaa@zju.edu.cn).}}

\begin{document}

\maketitle

\begin{abstract}
This paper presents a novel learning-based approach for online state estimation in flapping wing aerial vehicles (FWAVs). Leveraging low-cost Magnetic, Angular Rate, and Gravity (MARG) sensors, the proposed method effectively mitigates the adverse effects of flapping-induced oscillations that challenge conventional estimation techniques. 
By employing a divide-and-conquer strategy grounded in cycle-averaged aerodynamics, the framework decouples the slow-varying components from the high-frequency oscillatory components, thereby preserving critical transient behaviors while delivering an smooth internal state representation. 
The complete oscillatory state of FWAV can be reconstructed based on above two components, leading to substantial improvements in accurate state prediction. 
Experimental validations on an avian-inspired FWAV  demonstrate that the estimator enhances accuracy and smoothness, even under complex aerodynamic disturbances. 
These encouraging results highlight the potential of learning algorithms to overcome issues of flapping-wing induced oscillation dynamics.
\end{abstract}


\section{Introduction}

\IEEEPARstart{A}{utonomous} flapping flight, owing to its high
 maneuverability on par with birds and insects, offers promising potentials to tremendous applications \cite{shyy1999flapping}. 
Notably, avian-inspired flapping wing robots can accommodate additional onboard electronics and offer an extended operational range, thereby attracting increased attention in application-oriented research \cite{gerdes2014roboB,zufferey2021design,wenfu2022flight,chen2022novel}.
In order to enhance the survivability of Flapping Wing Aerial Vehicle (FWAV) in complex environments, it is imperative that their attitude estimation, as well as state prediction capabilities, are optimized for autonomous long-distance flight missions \cite{liu2022flying, wu2022long}.

The significance of attitude estimation lies in its ability to integrate noisy and constrained sensor data to produce a precise, real-time comprehension of the swift and unstable motions of a FWAV, thus facilitating safe and autonomous flight.
Although orientation estimation is vital for flapping wing autonomous flight, the community focus more on mechanisms and control techniques, and commonly borrow 
estimation method used in conventional robotic systems.
These off-the-shelf methods range from  Extended Kalman Filter (EKF) \cite{trawny2005indirect}, Mahony complementary filter \cite{mahony2008nonlinear} to gradient-based Madgwick filter \cite{madgwick2010efficient}.
According to our experience, these estimation algorithms can handle bird-like flapping wing flight with low accuracy requirement.
Nevertheless, researchers point out the notorious severe vibrations especially in accelerometer measurement
can adversely affect the overall performance of algorithms in specific applications \cite{tu2018realtime, verboom2015attitude}.
In early Delfly~II autonomous flight research \cite{verboom2015attitude}, a wingbeat-synchronized moving average filter was used to suppress accelerometer oscillations and estimate gravity for attitude determination. Later work \cite{armanini2017onboard} fused high-rate IMU data (512 Hz) with low-rate optical tracking (120 Hz) via an extended Kalman filter, enabling sub-cycle aerodynamic analysis without cycle-averaging. These studies reflect two contrasting strategies for handling flapping-induced periodic sensor patterns: to attenuate these oscillations or not to attenuate. 

\begin{figure}[t]
  \centering
  \includegraphics[width=3.2in]{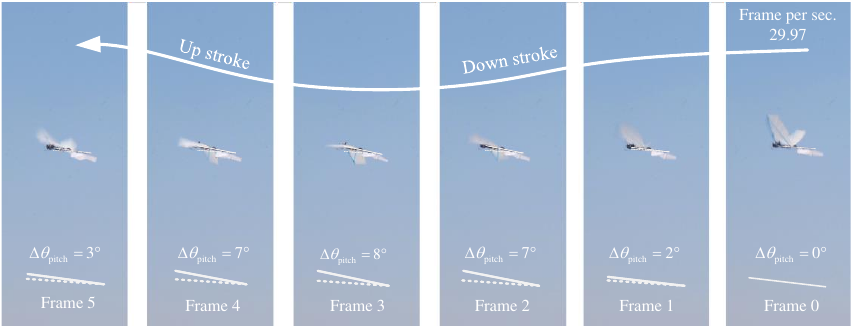}
  \caption{The FWAV in motion: The periodic motion of the flapping wings results in a non-negligible periodic pitch motion in the FWAV. }
  \label{Fig:Title}
\end{figure}

In avian-inspired robots with wingspans above one meter, the wing flapping frequency typically remains below 10 Hz, whereas inertial measurement units (IMUs) can sample at rates exceeding 500 Hz. Sufficiently high sampling rate enables the onboard IMU to capture the periodic wing motions, as demonstrated in \cite{grauer2009inertial}, even with low-cost MARG sensors. Although these sensors are usually mounted on the fuselage, they effectively register wing movements due to the transmitted fuselage motion (see Fig.~\ref{Fig:Title}). In \cite{tu2018realtime}, a model-based approach estimates oscillatory aerodynamic forces, positing that removing these forces from accelerometer data yields a more accurate gravitational acceleration estimate, however, we contend that the elimination should target only the specific force. Given the complex aerodynamic interactions, extracting periodic patterns via learning methods \cite{corban2023discovering,kim2024wing} is preferable over using stroke-angle models, particularly when such measurements are unavailable. By removing zero-mean periodic patterns, slowly varying measurements can be derived, thereby circumventing the typical delays in online cycle-averaging.

A critical component for achieving autonomous flapping-wing flight is the attitude control, which necessitates the production of specified wrenches. In the majority of relevant research, a straightforward and effective idea prevails across different scale flapping wing robots \cite{chen2022novel, chirarattananon2014adaptive, tu2020scale, qian2024toward}. 
This idea is to develop a map or a model which can translate the desired wrenches into specific wings and tail settings in a single flapping-wing period \cite{wenfu2022flight,chen2022novel,armanini2017onboard}.
This mapping generates parameters for the periodic movement, and these parameters should be preserved or only minimally adjusted over the duration of one entire flapping cycle. 
Simplified complexity often contributes to more stable control within FWAV research \cite{lu2024low}. 
Sudden aerodynamic changes can cause discrepancies between cycle-averaged wrenches and their expected values.
From an aerodynamic perspective, it is also crucial to eliminate fluctuating control inputs that may disrupt the airflow surrounding the FWAV, thereby impairing desired wrench generations.
The intuitive idea is to use cycle-averaged measurements as feedback signals.
However, the frequency of flapping in avian flight, like the developed FWAV shown in Fig.~\ref{Fig:Title}, is around 5~Hz. The latest-cycle averaging based attitude estimation, with a delay of approximately 0.1~s (half time of a cycle), is inadequate for achieving agile maneuver.

Modeling the dynamics is a crucial element for attaining online real-time planning and control.
The quasi-steady aerodynamics are collectively summarized in Sane and Dickinson's work \cite{sane2002aerodynamic}, which includes added mass effect, translational forces, rotational forces, and the wake-capture mechanisms.
Complementing the ideal map discussed earlier, the cycle-averaged model of quasi-steady aerodynamics is widely employed in application-focused research \cite{mcgill2021modeling, siqi2022modeling, sanchez2022simplified, wang2023modeling}. 
In their works, the wrenches are averaged over the extent of the wings, and assuming the fuselage oscillation plays a negligible role in the computation. 
However, wings flapping at approximately 5~Hz can substantially impact the dynamics of the fuselage. Additionally, the IMU measurement is mounted on the fuselage, where significant oscillations are also observed.

Building on the previous analysis, this work aims to develop a method for generating prompt, smooth measurements insulated from flapping-induced disturbances. 
These measurements enable reliable identification of parameters in an internal model, reflecting cognitive-level processing rather than low-level coarse perception. 
Combined with appropriate estimation algorithms, the internal model yields accurate internal state estimates while still preserving fast dynamics for faithful reconstruction of true oscillatory states.

In alignment with the objectives outlined in the motivation, the principal contributions of this study can be encapsulated in the following three aspects:
\begin{enumerate}
    \item We propose a novel learning-based method that can online estimate the flapping-induced oscillations in the proprioceptive sensing with low-cost MARG sensors. 
    \item  An internal model considering whole-body oscillations is developed based on cycle average, which captures the slow-varying behavior of the avian-inspired FWAV.
    \item By employing the divide-and-conquer approach to separately handle the slow-varying and oscillatory dynamics, and subsequently integrating these two components, an accurate comprehensive state estimation is achieved.
\end{enumerate}

The remainder of this paper is organized as follows.
In section II, we demonstrate the proposed online periodic pattern learning methods and the smooth state estimation method.
The internal model is developed in section III based on the cycle-averaged aerodynamics, as well as body oscillation information.
Then in section IV, the proposed algorithms and model are validated in real flight experiment of our avian-inspired FWAV.
Finally, we conclude this work in section V.

\section{Periodic Gaussian Process Learning}
Biological fliers accomplish smooth flight sensing by isolating oscillations caused by flapping-wing movements. 
Flapping-induced motions are largely isolated from the head, thereby reducing their impact on the vestibular system \cite{mcarthur2011state}.
Inspired by this, we propose a machine learning approach based on periodic Gaussian processes that autonomously learns and compensates for flapping disturbances in real time. 
By online periodic learning, the method extracts periodic components from acceleration and angular velocity signals, enabling smooth attitude estimation in FWAVs.

\subsection{Periodic Gaussian Process for Pattern Learning}
Gaussian Process Regression (GPR) is a non-parametric approach used in machine learning for regression tasks. 
It models the function generating the data as a sample from a Gaussian process.
This method is especially advantageous when the functional form is not known, and our aim is to deduce the periodic pattern from online flight data.

The aim of this learning process is to demonstrate that, given a phase value $\phi \in \mathcal S$ as input, the regression can effectively produce a significant estimation, such as determining the oscillatory component of the acceleration estimate.
The input values are denoted by the vector $\bm \Phi$, whereas the output values are indicated by the vector $\bm Y$.

To remove the non-zero prior mean from the periodic pattern, the following operation is implemented before feeding the data into GPR process:
\begin{align}
\bm Y_{\rm osc}=\bm Y-\cfrac{\sum_{i=1}^{{\rm dim}(\bm Y)}Y_i}{{\rm dim}(\bm Y)}
\end{align}
where $\bm Y_{\rm osc}$ is the novel output observation, 
\(Y_i\) represents the component in the vector $\bm Y$,
and \({\rm dim}(\bm Y)\) is the dimension of the output vector.

Given that the function we are modeling is recognized as nearly periodic, it is natural to use Fourier basis functions as the kernel to model periodicity \cite{Durrande-2016}.
Implemented as the feature vector function, the truncated Fourier basis is 
\begin{equation}
{\bm F} (\phi) = [ \sin(\phi) \; \cos(\phi) \; \cdots \; \sin(n \phi) \; \cos(n \phi) ]^\top
\end{equation}
where $\phi \in \mathcal S$ represents the phase of the periodic flapping-wing motion,
and $n \in \mathbb N^+$ is the highest harmonics of the phase.
Then we denote the feature space spanned by $\bm F$ is $\mathcal H_p$.
The reproducing kernel can be given as 
\begin{equation}
\label{eq:Rekernel1}
k_{p}(\phi_a, \phi_b)=\bm{F}^\top(\phi_a) {G}^{-1} \bm{F}(\phi_b)
\end{equation}
where $\phi_a$ and $\phi_b$ are two different phases, 
and $G$ is the gram matrix whose elements are shown as ${G}_{i, j}=\left\langle{F}_{i}, {F}_{j}\right\rangle_{\mathcal{H}}$.
Moreover, the inner product with respect to the space $\mathcal H$ can be given by
\begin{equation}
\left\langle{F}_{i}, {F}_{j}\right\rangle_{\mathcal{H}} = \frac{1}{2\pi} \int_{0}^{2 \pi}F_i(\psi)F_j(\psi){\rm d}\psi
\end{equation}
where $F_i$ represents the $i$th 
element of the feature vector function $\bm F$.
Consequently, the reproducing kernel given by \eqref{eq:Rekernel1} can be reformulated as 
\begin{equation}
\label{eq:Rekernel2}
k_{p}(\phi_a, \phi_b)= 2 \bm{F}^\top(\phi_a) \bm{F}(\phi_b)
\end{equation}
According to \cite{Quinonero-2005}, the joint distribution of the observed values and the function values at novel testing points is
\begin{equation}
\left[\begin{array}{c}\bm{Y}_{\rm osc} \\ {f}_{*}\end{array}\right] \sim \mathcal{N}\left( \bm 0,\left[\begin{array}{cc}K_p(\bm{\Phi}, \bm{\Phi})+\sigma_{n}^{2} I & {\bm K}_p\left(\bm{\Phi}, \phi_{*}\right) \\ {\bm K}_p\left(\bm{\Phi}, \phi_{*}\right)^\top & k_p\left(\phi_{*}, \phi_{*}\right)\end{array}\right]\right)
\end{equation}
where $\bm \Phi$ is the observed input vector, $\bm Y_{\rm osc}$ is the output observation vector,
$\phi_\star \in \mathcal S$ is the concerned phase location, and $f_\star \in \mathbb R$ is its corresponding output. 
The positive variable $\sigma_n \in \mathbb R^+$ represents the observation noises.
The covariance matrix can be constructed through the following manner:
\begin{equation}
K_{p,ij}=k_p(\phi_i,\phi_j)
\end{equation}
where $K_{p,ij} \in \mathbb R$ is the $i$-th row and $j$-th column of the matrix $K_p$.
Through investigating the conditional distribution, we can obtain the predictive equations for GPR:
\begin{subequations}
\label{eq:periodic_f_var}
\begin{align} 
\hat{f}_{*} & \triangleq \mathbb{E}\left[{f}_{*} \mid \mathbf{\Phi}, \bm{Y}, \phi_\star \right] \nonumber\\ & =\bm{K}_{p}^{\top}({\bm \Phi, \phi_\star})\left[{K}_{p}({\bm \Phi, \bm \Phi})+\sigma_{n}^{2} \bm{I}\right]^{-1} \bm{Y}_{\rm osc}, \label{eq:periodic_f_var_1}\\ 
\operatorname{var}\left(\hat{f}_{*}\right) & =k_p(\phi_\star, \phi_\star)\nonumber\\
 -\bm{K}_{p}^{\top}&({\bm \Phi, \phi_\star})\left[{K}_{p}({\bm \Phi, \bm \Phi})+\sigma_{n}^{2} \bm{I}\right]^{-1} \bm{K}_{p}({\bm \Phi, \phi_\star}) \label{eq:periodic_f_var_2}
\end{align}
\end{subequations}
where $\hat{f}_{*} \in \mathbb R$ and $\operatorname{var}\left(\hat{f}_{*}\right)$ are the estimation of the mean value and the variance, respectively.
Implementing these formulae, we can derive the output estimation and its associated uncertainty for an arbitrary phase position.

\begin{figure}[t]
  \centering
  \includegraphics[width=3.0in]{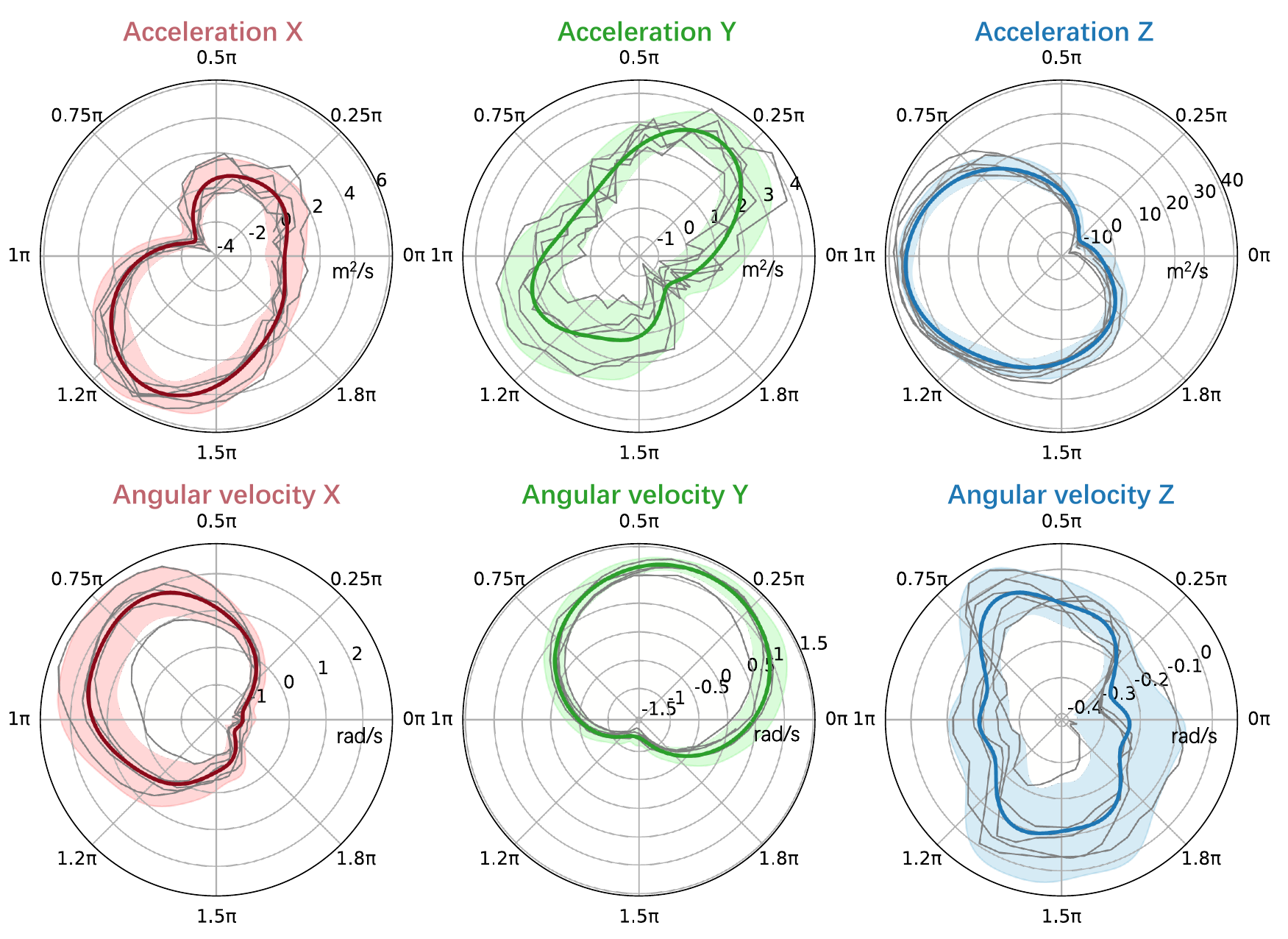}
  \caption{Polar plots illustrating periodic-GPR learning utilizing data that encompass four cycles.
  The unprocessed sensor data preserved for the purpose of periodic-GPR learning are depicted as gray lines.
  The mean values are illustrated with darker colors, while the areas corresponding to the two standard deviations are shaded in lighter tones. }
  \label{Fig:polarplot}
\end{figure}

As illustrated in Fig.~\ref{Fig:polarplot}, the periodic patterns of signals, notably the $Z$-component of acceleration and the $Y$-component of angular velocity, are effectively captured by periodic-GPR, which maintains uniformity across all phases, producing smooth learning results for further estimations.

\subsection{Period Length Estimation}
The Cooley--Tukey Fast Fourier Transform (FFT) algorithm efficiently computes the Discrete Fourier Transform (DFT) by recursively decomposing it into smaller DFTs using a \textit{divide-and-conquer} strategy \cite{Puschel-2008}.
High-order derivatives make both acceleration and angular velocity data sensitive to the periodic motion of flapping wings.
Real-time data streams contain these measurements with uneven timestamps. 
To address this, we interpolate the raw data to create an evenly spaced dataset that preserves its original frequency.

\begin{figure}[t]
  \centering
  \includegraphics[width=3.0in]{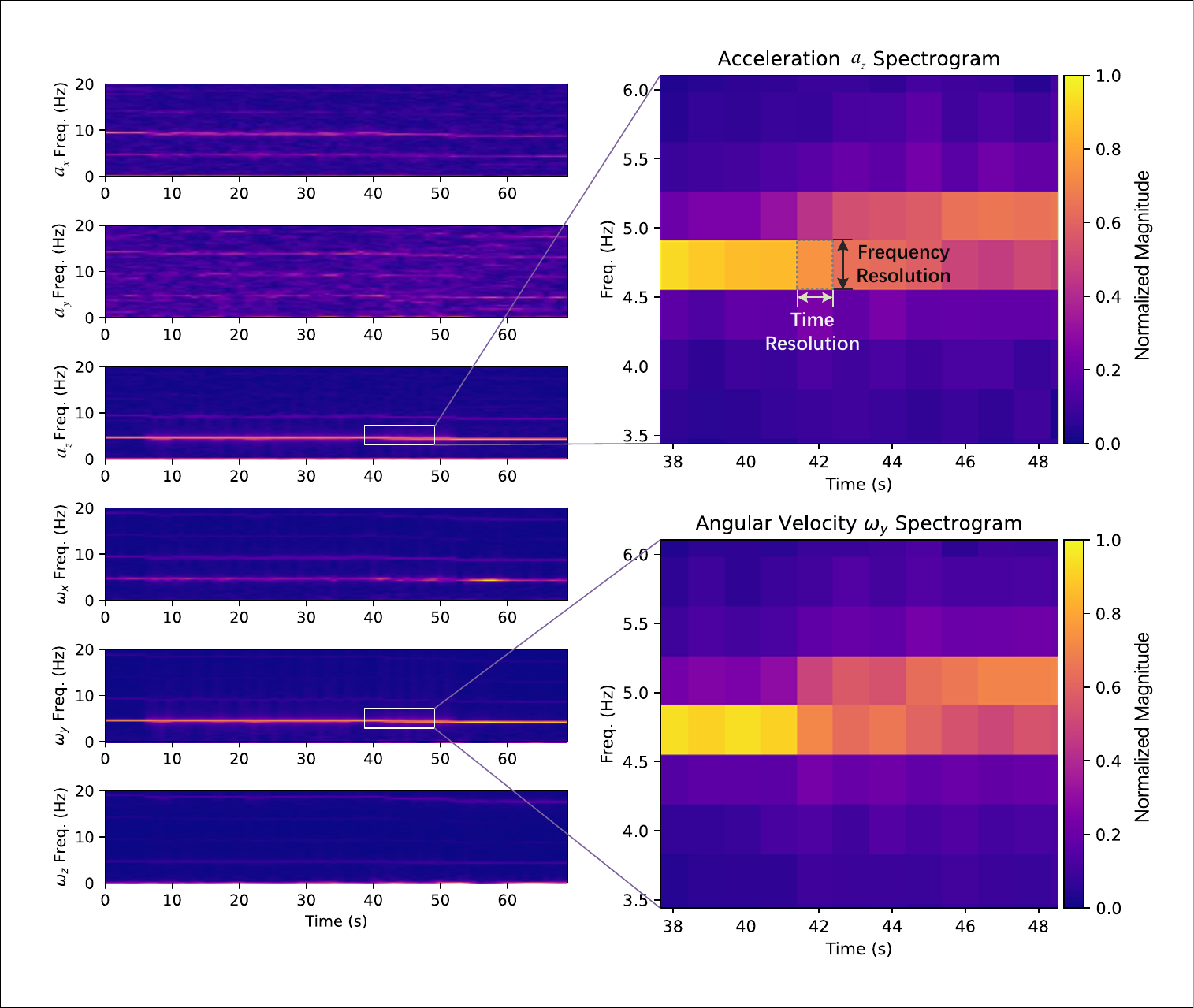}
  \caption{Spectrograms over time generated by sliding-window Cooley-Tukey FFT alogorithm with real flight data. The signals of acceleration and angular velocity are employed for frequency estimation due to their superior update rates compared to all other available signals.}
  \label{Fig:Spec}
\end{figure}

The sliding-window FFT results from real-flight data are shown in Fig.\,\ref{Fig:Spec}.
The frequency data can be extracted from the $Z$-component of acceleration $a_z$ and the $Y$-component of angular velocity $\omega_y$ signals.
The right sub-figures demonstrate that the frequencies extracted from both signals are closely aligned, indicating that the frequency extraction process is relatively stable.
In DFT analysis, improving frequency resolution by increasing the observation window compromises time resolution, as it limits the ability to capture rapid temporal variations. Conversely, enhancing time resolution reduces frequency resolution, making it more difficult to distinguish closely spaced frequencies.
Nevertheless,  at a specific moment $t$ in time, employing a Gaussian distribution to model the frequency can offer an analytical representation of the frequency, 
which has the mean value $f_{\rm freq}$ and the variance $\sigma^2_{\rm freq}$, as a result of integrating frequency observations from $a_z$ and $\omega_y$.

\subsection{Signals Smoothing by k-means on $\mathcal S$}
Sphere k-means is particularly effective for managing phase-based data, such as angular and directional data, since it operates under the premise that data points are situated on the surface of a unit sphere \cite{hornik-2012}.

In this context, the cosine similarity can be expressed as 
\begin{equation}
d_{\cos}( \phi_1,\phi_2) = 1 - \cos(\frac{\phi_1 - \phi_2}{2})
\end{equation}
where $\phi_1$ and $\phi_2$ are two distinct phase angle.
The k-means cluster algorithm can be transformed into the optimization problem $\min_{C_\phi}\mathcal C ( C_\phi )$. 
The objective function for this optimization problem is formulated as
\begin{subequations}
\label{eq:phase_cluster_cost}
\begin{align}
&\mathcal C ( C_\phi )=\sum_{\phi_{d}\in D_{\phi}}\sum_{\phi_{c}\in C_{\phi}}\delta(l_{C_{\phi}}(\phi_{d}),\phi_{c})d_{\cos}( \phi_{d},\phi_{c} )\\
&l_{C_{\phi}}(\phi) = \underset{\phi_c \in C_{\phi}}{\rm arg\,min} \,d_{\cos}(\phi,\phi_c)\\
&\delta(\phi_1,\phi_2) = 
\begin{cases}
    1 & \text{if } \phi_1 = \phi_2 \\
    0 & \text{if } \phi_1 \neq \phi_2
\end{cases}
\end{align}
\end{subequations}
where $D_\phi$ is the set of phases corresponding to the overall data stored in memory, and $C_\phi$ is the set of cluster centroids based on the data in memory. 
The function $l_{C_{\phi}}(\phi)$ identifies the closest cluster centroid relative to $\phi$.
The Kronecker delta function $\delta$ is employed to constrain the distance accumulation within each category.
In order to improve algorithmic efficiency, the optimization-based k-means is initially employed to produce more evenly distributed cluster centers, which is subsequently followed by the application of the traditional k-means algorithm, owing to its higher computational efficiency.

After k-means clustering, each data point is assigned a cluster label. Evaluation is then performed by averaging values within each cluster, producing both mean signals and phase angles for periodic-GPR learning. The observation variance per cluster is also directly computed and used as $\sigma_n^2$ in \eqref{eq:periodic_f_var_1}. As shown in Fig.~\ref{Fig:polarplot}, with only 16 cluster centers, uncertainties in the periodic patterns are mainly attributed to observation noise rather than distance from cluster centers. 

\subsection{Proprioceptive Learning \& Estimation Pipeline}
The aforementioned algorithm necessitates an initial phase, indicated as $\phi_0$.
Utilizing the B-spline frequency estimation, the instantaneous phase angular velocity can consistently be determined as $2 \pi / \hat{f}_{\rm freq}$.
The phase angle of the flapping-wing motion is then updated by
\begin{subequations}
\label{eq:phase_pred}
\begin{align}
\phi(t_{\rm curr.})&=\phi(t_{\rm last})+2 \pi (t_{\rm curr.} - t_{\rm last})/ \hat{f}_{\rm freq}\\
\phi(t_0)&=\phi_0
\end{align}
\end{subequations}
where $t_{\rm curr.} \in \mathbb R$ is the current time instance,
and $t_{\rm last} \in \mathbb R$ is the preceding time instance. 
To improve the efficiency of periodic learning, phase correction is utilized and accomplished through the application of cross-correlation:
\begin{subequations}
\label{eq:phase_update}
\begin{align}
R_{\rm cc}(k)&= \sum_{i = - \lceil{f_{\rm s}} /{ f_{\rm freq}}\rceil } ^{\lceil{f_{\rm s}} /{ f_{\rm freq}}\rceil} {\bm s}(i) \top \hat  {\bm s}(i+k)\\
\phi_{\rm d} &= \frac{2 \pi  f_{\rm freq}}{f_{\rm s}} \underset{k \in [-{\lceil{f_{\rm s}} /{ f_{\rm freq}}\rceil} , {\lceil{f_{\rm s}} /{ f_{\rm freq}}\rceil} ]}{\rm arg\,max}  R_{\rm cc}(k) \\
\phi'(t_{\rm curr.}) &= \phi(t_{\rm curr.})  + k_{\rm cc} \min (\max (\phi_{\rm d}, \underline\phi_{\rm d}), \overline\phi_{\rm d})
\end{align}
where $R_{\rm cc}(k) \in \mathbb Z \to \mathbb R$ is the cross-correlation with difference of $k$, with $k\in \mathbb Z$ being the data sequence number, 
 $\lceil \star \rceil \in \mathbb R \to \mathbb Z$ rounds up $\star$ to the nearest integer,
 the selected signal vector
 $\bm s = [ a_z \; \omega_y]^\top \in \mathbb R^2 $ collects the signals with distinctive frequencies, meanwhile,
  $\hat{\bm s} \in \mathbb R^2 $ is its corresponding estimation from GPR.
Moreover, $f_s \in \mathbb R$ is the sampling frequency of the sensors.
Furthermore, $\phi_{\rm d} \in \mathbb R$ signifies the phase difference. Concurrently, $\underline \phi_{\rm d} \in \mathbb R^-$ and $\overline \phi_{\rm d} \in \mathbb R^+$ indicate its lower and upper bounds, respectively.
And the constant $k_{\rm cc} \in \mathbb R^+$ is the correction update rate.
Finally, we obtain the novel phase angle estimation $\phi' \in \mathcal S$.

\end{subequations}
Periodic stimuli patterns recognition involves analyzing acceleration, angular velocity, geomagnetic sensing, and translational velocity within the wind frame.
These signals are categorized into distinct groups based on their update rates to facilitate their data selection and learning of periodic patterns.

In the explanation of this subsection, we utilize periodic acceleration learning as an illustrative example.
Since the phase estimation \eqref{eq:phase_pred} and \eqref{eq:phase_update} can be obtained at any time instance, a data tuple as $\lbrace \bm a_\star, \phi_\star \rbrace$ becomes available immediately upon receiving acceleration data.
These data tuples are subsequently stored in a memory stack, where older entries are replaced by new ones once the stack reaches its capacity.
 Thus, the data are selected by k-means method as illustrated in \eqref{eq:phase_cluster_cost}.

\begin{figure*}[t]
  \centering
  \includegraphics[width=5.0in]{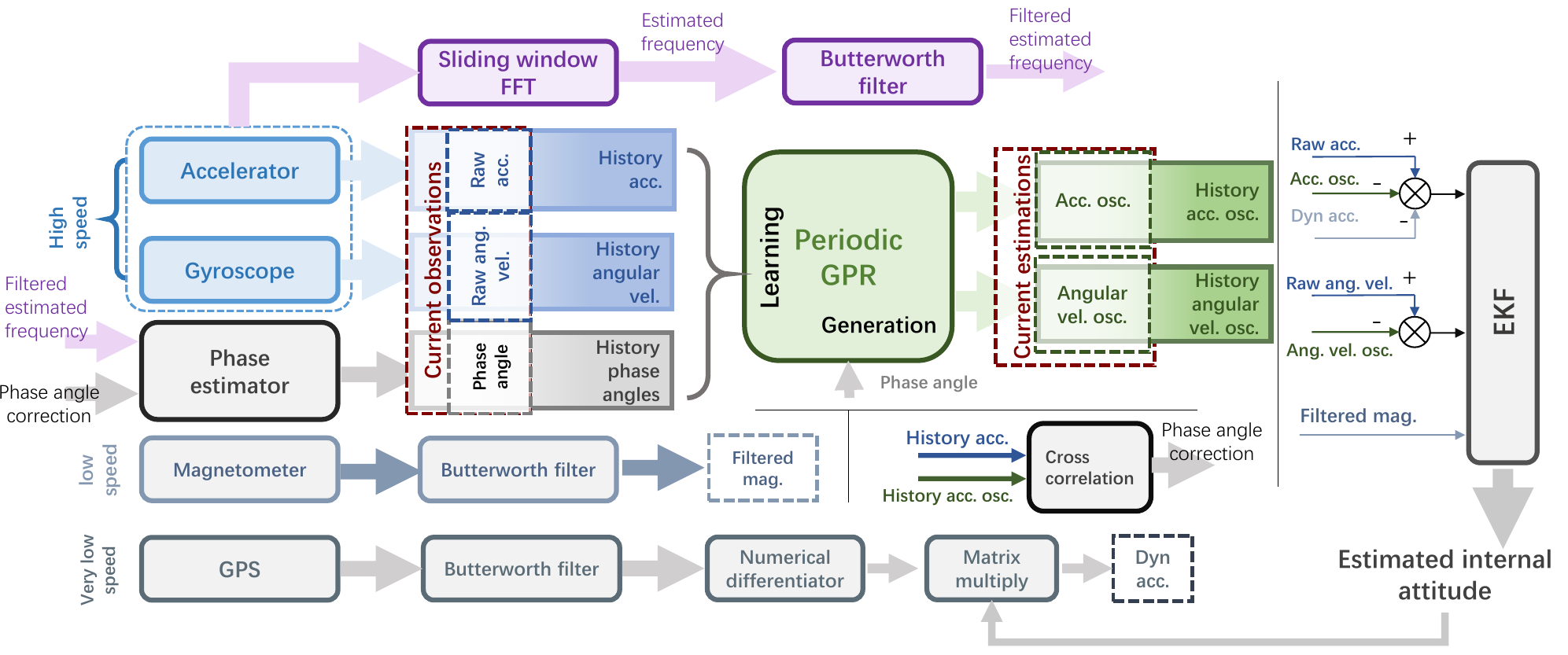}
  \caption{Framework for implementing periodic Gaussian process learning in the estimation of internal attitude states. }
  \label{Fig:overall}
\end{figure*}

The comprehensive state estimation method is illustrated in Fig.~\ref{Fig:overall}. The phase estimation results, along with the rapidly updated raw data exhibiting oscillations due to the flapping-wing motion—specifically the acceleration and angular velocity—are inputs for the periodic GPR learning. 
Upon provision of a phase angle, the periodic GPR process outputs the respective oscillation components, which are consistently estimated in real-time. 
These oscillation-free signals are subsequently processed by an EKF to estimate attitude unaffected by flapping-wing-induced oscillations.

 The selected data tuples are then fed into GPR process with periodic kernels.
 Consequently, when a phase input is given, it is capable to estimate the acceleration associated with that phase by implementing \eqref{eq:periodic_f_var}. 
 With the raw acceleration observation $\bm a$ and the estimated acceleration oscillation $\hat {\bm a}_{\rm osc}$ in hand,
 the internal acceleration can be obtained as $\bm a - \hat {\bm a}_{\rm osc}$.
 The angular velocity data undergoes the same operation, whereas the geomagnetic vector data and Global Positioning System (GPS) data do not, due to their slow variation, lower sampling rate, and reduced susceptibility to oscillation disturbances.
Subsequently, the internal signals are processed using the standard EKF method \cite{Sabatini-2011}, 
enabling the estimation of the internal attitude estimation of the robot.
The dynamic acceleration is estimated using feedback from the GPS system. 
The gravitational acceleration is then determined by subtracting the dynamic acceleration from the total acceleration measurement.

\section{Forward Slow-varying Internal Model}
An internal model in the sensorimotor system is a neural mechanism that predicts the sensory outcomes of actions to enable accurate and adaptive motor control \cite{wolpert1995internal}.
Inspired by this concept, we develop an internal model for FWAV translational dynamics using averaging theory. 
The derivation focuses on longitudinal dynamics based on Newton-Euler equations, while the complete model captures full 3D motion by combining cycle-averaged aerodynamic coefficients from offline measurements and online learning with configuration-specific parameters.

\begin{figure}[t]
  \centering
  \includegraphics[width=2.8in]{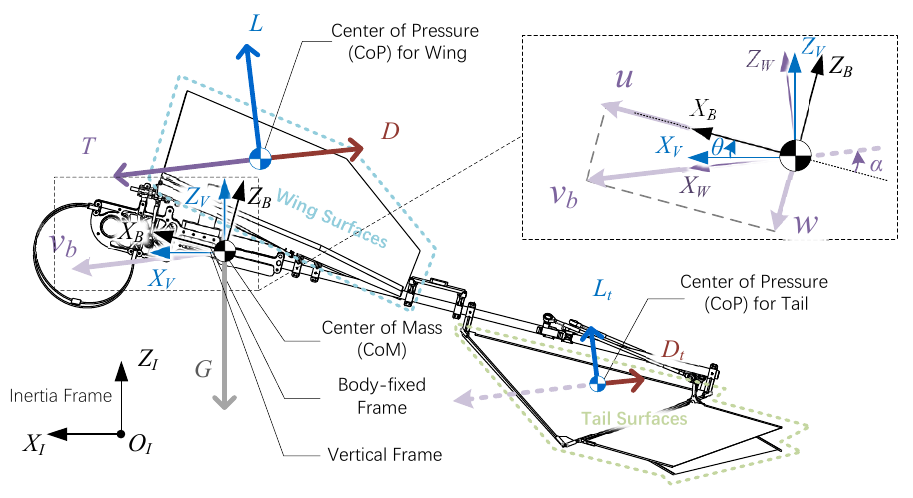}
  \caption{The longitudinal reference framework of the FWAV along with forces exerted upon it. 
  The longitudinal inertia frame is denoted by $X_{I}$-$Z_{I}$, whereas the longitudinal body-fixed frame is represented by $X_{B}$-$Z_{B}$. 
  Assuming the FWAV can execute banked turns without any lateral slip, we define a vertical frame, \(X_V\)–\(Z_V\), based individually on its yaw rotation. 
  The wind frame, \(X_W\)–\(Z_W\), is defined based on the translational velocity.
  The deflection angles of the ruddervators on the inverted-V tail are anticipated to cause minimal variation in lift on the tail surfaces.}
  \label{Fig:dynamic_model}
\end{figure}

\subsection{Longitudinal Newton-Euler Dynamic Equations}
According to \cite{tal2021global}, the lateral force is considered negligible due to the relatively small lateral surface area of the FWAV and the absence of a vertical tail. 
We only discuss longitudinal dynamics in the following sections.
The longitudinal translational and rotational dynamics of the robot are described utilizing the Newton-Euler equations.
The related frames and aerodynamic forces during flight are clearly illustrated in Fig~\ref{Fig:dynamic_model}. Assuming that thrust line passes through the center of gravity, equations for velocity along $X_{B}$ and $Z_{B}$, denoted as $u$ and $w$, pitch angle and angular rate, denoted as $\theta$ and $q$, can be given by
\begin{subequations}
\label{eq:_dot_s}
\begin{align}
\label{eq:u_dot_}
\frac{{\rm d} u}{{\rm d}t}&= \frac{(L+L_{t})\sin{\alpha}+(T-D-D_{t})\cos{\alpha}}{m}\nonumber\\&~~~~+
g\sin{\theta}- qw\\
\label{eq:w_dot_}
\frac{{\rm d} w}{{\rm d}t}&= \frac{(L+L_{t})\cos{\alpha}-(T-D-D_{t})\sin{\alpha}}{m}\nonumber\\&~~~~-g\cos{\theta}+qu\\
\label{eq:theta_dot_}
\frac{{\rm d} {\theta}}{{\rm d}t}&= {q}\\
\label{eq:q_dot_}
\frac{{\rm d} q}{{\rm d}t}&= \frac{(Ll_{w}+L_{t}l_{t})\cos{\alpha}}{I_{p}}
\end{align}
\end{subequations}
where $\alpha \in \mathbb R$ is the angle of attack (AoA). $m \in \mathbb R$ and $I_{p} \in \mathbb R$ represent the robot mass and the moment of inertia around $Y_{b}$. 
Furthermore, the symbols $l_{w} \in \mathbb R^+$ and $l_{t} \in \mathbb R^+$ denote the moment arms from the center of pressure (CoP) of the wing and tail to the center of mass (CoM) within the longitudinal plane, respectively. 
Let \( L \in \mathbb{R} \), \( T \in \mathbb{R} \), and \( D \in \mathbb{R} \) represent the lift, thrust, and drag produced by the wing, respectively. 
Meanwhile, \( L_{t} \in \mathbb{R} \) and \( D_{t} \in \mathbb{R} \) denote the lift and drag generated by the tail, respectively.
Finally, the gravity acceleration is denoted by $g \in \mathbb R^+$.

The path followed by the wing tip as a function of time $t \in \mathbb{R}$ is described by the equation $h(t) = h_0 \sin(2\pi ft)$., where $f$ represents the frequency of flapping and $h_0$ denotes the amplitude of flapping associated with the reference chord.
The flapping phase $\phi \in \mathbb{R}$ is defined by the equation $\phi = 2\pi ft$.
Aerodynamic forces are represented through aerodynamic coefficients denoted as $C_{\star}$, where $\star$ may correspond to $L, T, D$ and $\star_{t}$ correspond to $L_{t}, D_{t}$.
The particular coefficient is given by $C_{\star} = ({\star}+{\star_{t}})/{QS}$, in which $Q\in \mathbb{R}$ denotes the dynamic pressure, and $S$ represents the wing area.
The aerodynamic coefficients we calculate are grounded in the framework established by Ollero et al. \cite{rodriguez-2022}. 
As stated by \cite{sane2002aerodynamic}, lift is comprised of translation force, rotational force, wake capture force, and inertia force.
Given that they oscillate with the same period, we simplify by representing the phase difference for the lift coefficient oscillation angle as $\phi_{L}\in \mathbb{R}$, and the amplitude is indicated as $C_{L, \rm osc}\in \mathbb{R}^+$. The lift coefficient $C_{L}\in \mathbb{R}^+$ is 
\begin{equation}
\label{eq:CL}
C_{L} =  C_{L,\alpha} \bar{\alpha} + C_{L, \rm osc} \sin{(\phi + \phi_{L})}
\end{equation}
where $C_{L,\alpha}\in \mathbb{R}^+$ is lift curve slope, $\bar{\alpha} \in \mathbb{R}$ is periodic average AoA. $V_{b}\in \mathbb{R}$ is body velocity relative to wind. 
We define the periodic oscillation of AoA as $\tilde{\alpha}_{osc} =  \alpha_{\rm osc}\sin{(\phi + \phi_{\alpha})}$, where $\alpha_{\rm osc}\in \mathbb{R}$ is oscillation amplitude of AoA and $\phi_{\alpha}\in \mathbb{R}$ is phase difference between oscillation phase and $\phi$, which leads thrust coefficient $C_{T}\in \mathbb{R}$ to
\begin{align}
\label{eq:CT}
C_{T} = & {C_{Th}}(k_{\rm red.} h_{0})^{2}\sin(2\pi ft)[G_{1}(k_{\rm red.})\sin(2\pi ft) \nonumber \\
& - F_{1}(k_{\rm red.})\cos(2\pi ft)] + (\bar{\alpha} + \tilde{\alpha}_{osc}) C_{L}
\end{align}
where $C_{Th}\in \mathbb{R}^+$ represents the coefficient defined in \cite{sanchez-2022}, $G_{1}(k_{\rm red.})\in \mathbb{R}$ and $F_{1}(k_{\rm red.})\in \mathbb{R}$ are respectively the imaginary and real parts of the Theodorsen's function $C_{1}(k)$ in \cite{fernandez-2016}. 
Moreover, $k_{\rm red.}\in \mathbb{R}$ is reduced frequency defined as $k_{\rm red.}=2\pi f/V_{b}$.

Drag coefficient $C_{D}\in \mathbb{R}^+$ contains zero-lift drag coefficient $C_{D0}\in \mathbb{R}^+$ and induced drag coefficient $C_{Di}\in \mathbb{R}^+$. $C_{D0}$ corresponds to the friction and manifests as a constant, simultaneously, $C_{Di}$ is induced by $C_{L}$, inversely proportional to aspect ratio of the wing. The inversely proportional coefficient is denoted as $A$. Therefore, $C_{D}$ is demonstrated as
\begin{equation}
\label{eq:CD}
C_{D}=C_{D0}+C_{Di} \approx C_{D0}+\frac{C_{L}^{2}}{A}
\end{equation}

\subsection{Internal Model by Averaging Theory}
In this subsection, we derive internal aerodynamic coefficients by cycle averaging magnitude of aerodynamic coefficients projected on cycle average velocity framework in a flapping period:
\begin{subequations}
\label{eq:C_inte}
\begin{align}
\label{eq:C_L_inte}
\bar C_{L}&=\frac{1}{2\pi}\int_{0}^{{2\pi}}(C_{L}\cos{\tilde{\theta}_{V,{\rm osc}}}-(C_{T}-C_{D})\sin{\tilde{\theta}_{V,{\rm osc}}}){\rm d}\phi \\
\label{eq:C_T_inte}
\bar C_{T}&=\frac{1}{2\pi} \int_{0}^{{2\pi}}(C_{T}\cos{\tilde{\theta}_{V,{\rm osc}}}+C_{L}\sin{\tilde{\theta}_{V,{\rm osc}}}){\rm d}\phi \\
\label{eq:C_D_inte}
\bar C_{D}&=\frac{1}{2\pi}\int_{0}^{{2\pi}}C_{D}\cos{\tilde{\theta}_{V,{\rm osc}}}{\rm d}\phi
\end{align}
\end{subequations}
where $\tilde{\theta}_{V,{\rm osc}} = \tilde{\alpha}_{\rm osc}+\tilde{\theta}_{\rm osc}$ and  $\tilde{\theta}_{\rm osc}$ denote periodic oscillation in velocity direction and pitch angle.

Let $\tilde{u}_{\text{osc}}\in \mathbb{R}$ and $\tilde{w}_{\text{osc}}\in \mathbb{R}$ represent the periodic oscillations in $u$ and $w$, respectively, and $\tilde{q}_{\text{osc}}$ denotes the corresponding periodic oscillation in pitch angular velocity.
In order to achieve a balance between computational efficiency and accuracy, we neglect the insignificant term $\tilde{{u}}_{\text{osc}}$ and assume that the periodic oscillations in $\tilde{{w}}_{\text{osc}}$ and $\tilde{{q}}_{\text{osc}}$ predominantly stem from fluctuations in the lift coefficient within the wrenches.
Under small AoA assumption, velocity along $\rm X$-axis of the body-fixed frame $u \approx V_{b}$, and we subsequently obtain $\tilde{\alpha}_{{\rm osc}}=\arctan2(-\frac{\tilde{w}_{{\rm osc}}}{u}) \approx -\frac{\tilde{w}_{{\rm osc}}}{V_{b}}$
Rotational force generated by the tail is insignificant \cite{sanchez-2022} and neglect periodic oscillation of torque generated by the tail. The equation (\ref{eq:w_dot_}-\ref{eq:q_dot_}) are transformed into
\begin{subequations}
\label{eq:angle_dot_s}
\begin{align}
\label{eq:alpha_dot}
\frac{{\rm d} \tilde{\alpha}_{{\rm osc}}}{{\rm d}t}&= -{C}_{L, {\rm osc} }\sin{(\phi + \phi_{L})}\frac{QS}{mV_{b}} - \tilde{q}_{{\rm osc}}\\
\label{eq:theta_dot}
\frac{{\rm d} \tilde{\theta}_{{\rm osc}}}{{\rm d}t}&= {\tilde{q}_{{\rm osc}}}\\
\label{eq:q_dot}
\frac{{\rm d}\tilde{q}_{{\rm osc}}}{{\rm d}t}&= {C}_{L,{\rm osc}}\sin{(\phi + \phi_{L})}\frac{QSl_{w}}{I_{p}}
\end{align}
\end{subequations}

Denote the oscillation amplitude of $\tilde{q}_{\rm osc}$ as $q_{\rm osc}\in \mathbb{R}^+$. Solving \eqref{eq:alpha_dot} and \eqref{eq:q_dot}, phase difference $\phi_{d}\in \mathbb{R}$ between $\phi_{\alpha}$ and $\phi_{L}$ is provided by
\begin{align}
\label{eq:phase_difference_alpha}
\cos{\phi_{d}}=\frac{q_{{\rm osc}}}{\sqrt{(q_{{\rm osc}})^{2}+(\frac{C_{L,{\rm osc}}QS}{mV_{b}})^{2}}}
\end{align}

Substitute (\ref{eq:CL}-\ref{eq:CD}), and (\ref{eq:phase_difference_alpha}) into (\ref{eq:C_L_inte}-\ref{eq:C_D_inte}) and neglecting small variables, internal aerodynamic coefficients is 
\begin{align}
\label{eq:C_L_inte_}
\bar C_{L}&= C_{L,\alpha} \bar \alpha\\
\label{eq:C_T_inte_}
\bar C_{T}&= \frac{1}{2}C_{Th}(k_{\rm red.} h_{0})^{2}G_{1}(k) + \bar \alpha^{2}C_{L,\alpha}+\frac{C_{L,{\rm osc}}q_{{\rm osc}}}{4\pi f}\\
\label{eq:C_D_inte_}
\bar C_{D}&=C_{D0}+\frac{1}{A}(C_{L,\alpha}^{2}\bar \alpha^{2}+\frac{1}{2}C_{L,{\rm osc}}^{2})
\end{align}

During flight, $C_{L,{\rm osc}}$ and $q_{{\rm osc}}$ are estimated online with the help of the method given in the last section. 
In contrast, the remaining coefficients are completely determined by the robot aerodynamic configuration.

Finally, the overall translational
 internal model in three-dimensional space can be expressed by
 \begin{subequations}
\label{eq:internal_model}
\begin{align}
\label{eq:p_dot}
\frac{{\rm d}{\bar {\bm p}}}{{\rm d}t}&=\bar {\bm v}\\
\label{eq:v_dot}
\frac{{\rm d} \bar{\bm v}}{{\rm d}t}&=-g\bm e_{3} + R_{B}^{I}R_{W}^{B}\bm f^{W}
\end{align}
\end{subequations}
The aerodynamic force in the wind frame ${\bm f}^W \in \mathbb R^3$ is expressed as ${\bm f}^W=QS\cdot\left [ \bar C_T- \bar C_D \quad 0 \quad \bar C_L\right ]^\top$.
Let $\bar {\bm p}\in \mathbb R^3$ and $\bar {\bm v}\in \mathbb R^3$ denote the internal position and velocity vectors, respectively. The matrix $R_{B}^{I}\in \mathrm{SO}(3)$ represents the rotation from the body-fixed frame to the inertial frame, while $R_{W}^{B}\in \mathrm{SO}(3)$ denotes the rotation matrix from the wind frame to the body-fixed frame.
Moreover, $\bm e_{3}=[0\quad0\quad1]^\top$ represents the $Z$-axis of the inertia frame.

\section{Experiments \& Analyses}
\begin{figure}[t]
  \centering
  \includegraphics[width=3.0in]{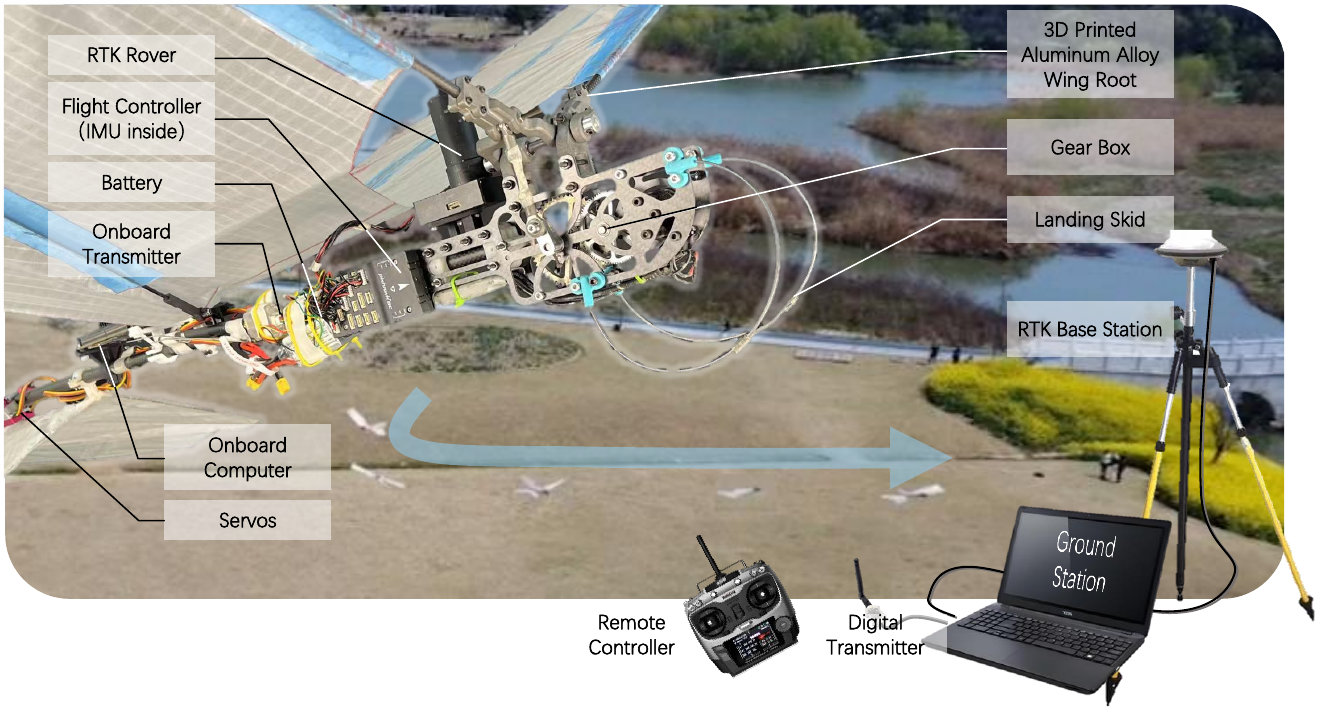}
  \caption{Real flight experiment settings: The FWAV is remotely controlled, supported by an attitude control realized onboard.
High-precision positioning is achieved by fusing Real-Time Kinematic (RTK) rover signals onboard the FWAV with data transmitted from the RTK base station.}
  \label{Fig:Settings}
\end{figure}

Real flight experiments for the developed FWAV is carried out at the Zijingang Campus, Zhejiang University \footnote{See video at https://www.youtube.com/watch?v=lavNlu5x11Q and codes https://github.com/XJX-UAV/Flapping-wing-State-Estimation.}. 
As explicitly illustrated in the setting demonstration, Fig.~\ref{Fig:Settings}, the FWAV is equipped with a DC brushless motor, an Electric Speed Controller (ESC), a Holybro Pixhawk 6C flight controller, an NVIDIA Jetson Orin NX onboard computer, an 800 mAh 4S LiPo battery, and two servos for controlling the ruddervators. Moreover, the FWAV possesses a wingspan measuring 1.7~m and a mass of 1100~g.
The online estimation algorithm is implemented using C++ and operates in conjunction with the Robot Operating System (ROS) on a Linux Ubuntu platform.

The desired attitude is manually input and subsequently provided to the attitude controller. The attitude control is executed onboard by the flight controller.
Meanwhile, the MARG sensor signals are transmitted from the flight controller to the onboard computer with corresponding timestamps. 
The acceleration and angular velocity data are refreshed at an approximate frequency of 170.8 Hz, while the magnetic data are updated at a frequency of roughly 13.4 Hz. Additionally, the GPS data are refreshed at an approximate frequency of 8.7 Hz. Second-order Butterworth filters are utilized to standardize the data frequencies to 200.0 Hz.

\subsection{Periodic GPR Learning}
\begin{figure}[t]
  \centering
  \includegraphics[width=2.8in]{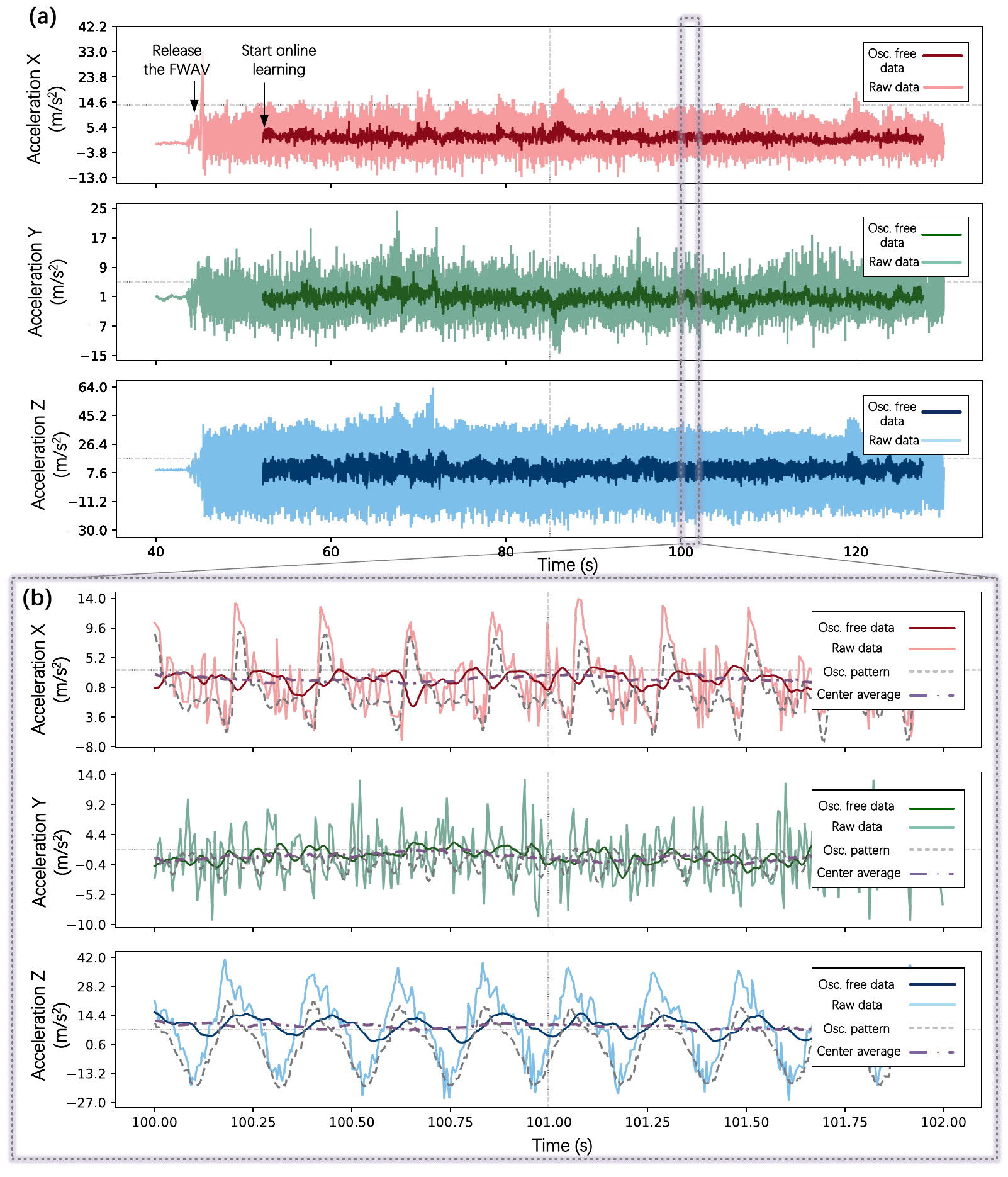}
  \caption{Acceleration measurements in real flapping-wing flight: 
  Dark solid lines denote oscillation-free data, and light solid lines represent raw data. The violet dot-dashed line shows cycle-averaged data, available only offline, used to approximate the ground truth.
  The learned oscillatory patterns are represented by gray dashed lines. }
  \label{Fig:Acc_res}
\end{figure}

The periodic GPR learning algorithm functions in real-time and is implemented onboard.
Meanwhile, the FFT algorithm is performed based on the $Z$-component of acceleration, $a_z$, and the $Y$-component of angular velocity, $\omega_y$, each with a length of 512 data points. 
Given that the flapping-wing frequency is approximately 5.0~Hz, frequencies less than 1.0~Hz or exceeding 8.0~Hz are excluded from the FFT intensity evaluation.
The cross-correlation is performed based on one flapping-wing cycle data. 
Furthermore, in the k-means algorithm utilized, a total of 16 cluster centers are implemented, and these are computed in real-time. 
Given the experimental settings described above, the acceleration results of the online periodic pattern learning are ultimately depicted in Fig.~\ref{Fig:Acc_res}.
To demonstrate the robustness of the proposed algorithm, the experiment is conducted under windy conditions, with wind speeds above 5.0 m/s.

\begin{figure}[t]
  \centering
  \includegraphics[width=3.0in]{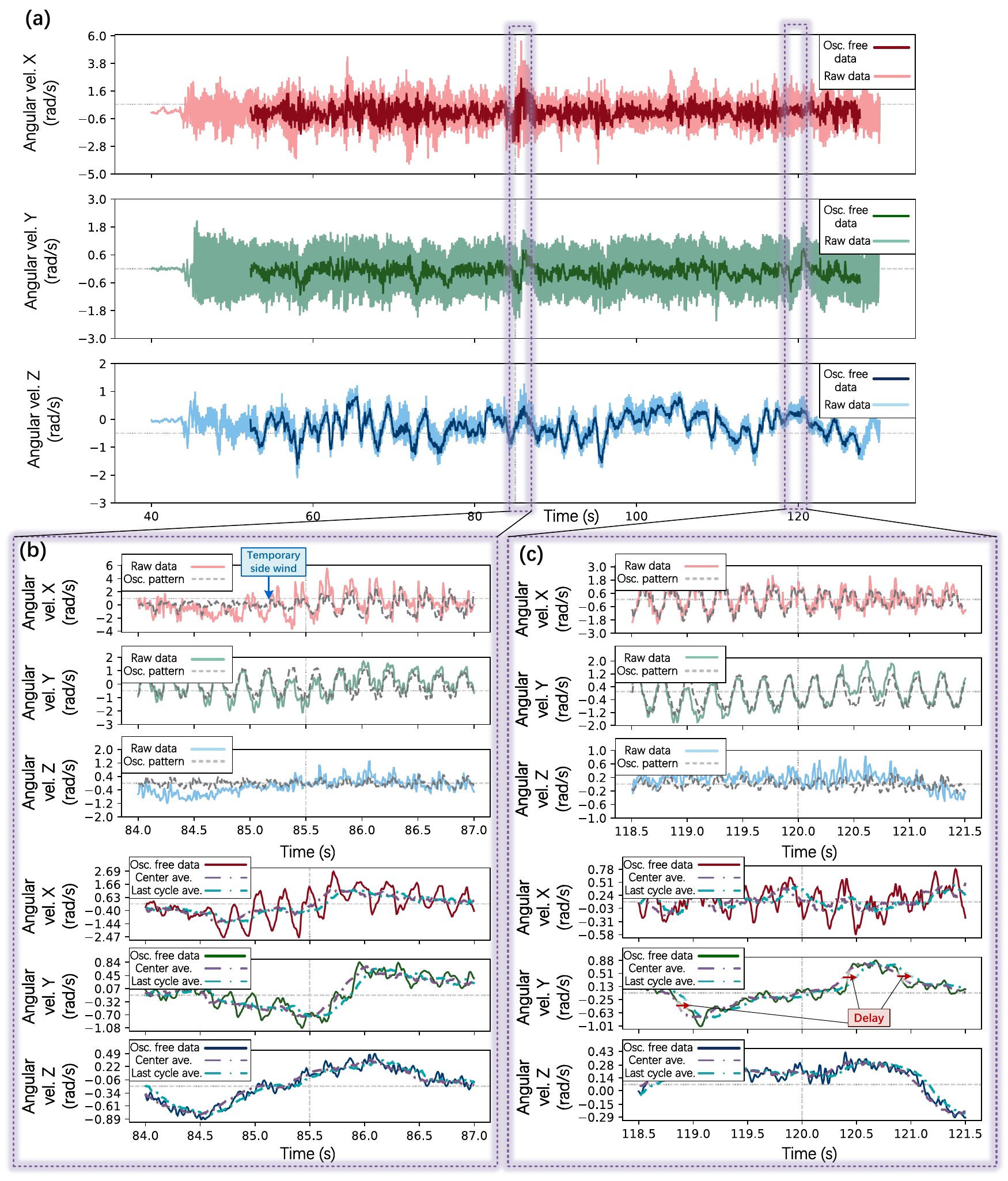}
  \caption{Angular velocity measurements in real flapping-wing flight: The curves graphical representations are assigned similarly to those in Fig.~\ref{Fig:Acc_res}. 
  The average over last cycle is illustrated as teal dashed lines.}
  \label{Fig:Agv_res}
\end{figure}
As demonstrated in Fig.~\ref{Fig:Acc_res}-(a), 
the proposed algorithm effectively remove oscillatory measurements, resulting in signals with reduced oscillatory characteristics that are more amenable to subsequent internal attitude estimation.
In the zoom-in view presented in Fig.~\ref{Fig:Acc_res}-(b), empirical analysis reveals that, with the application of FFT and cross-correlation corrections, the periodic pattern derived through online learning demonstrates a pronounced alignment with the observable trends in the raw data. 
For practical purposes, high-cutoff frequency filtering is applied to the raw observation values minus the learned oscillation patterns, resulting in acceleration data free of oscillations.
Moreover, it can be observed that the oscillation-free acceleration data remains in close proximity to the center averaged data via a sliding window with a length equivalent to one flapping cycle.
The center averaged data serves as an approximation of the ground truth. However, its online computation is impractical because future data is not available.
\begin{figure}[t]
  \centering
  \includegraphics[width=2.8in]{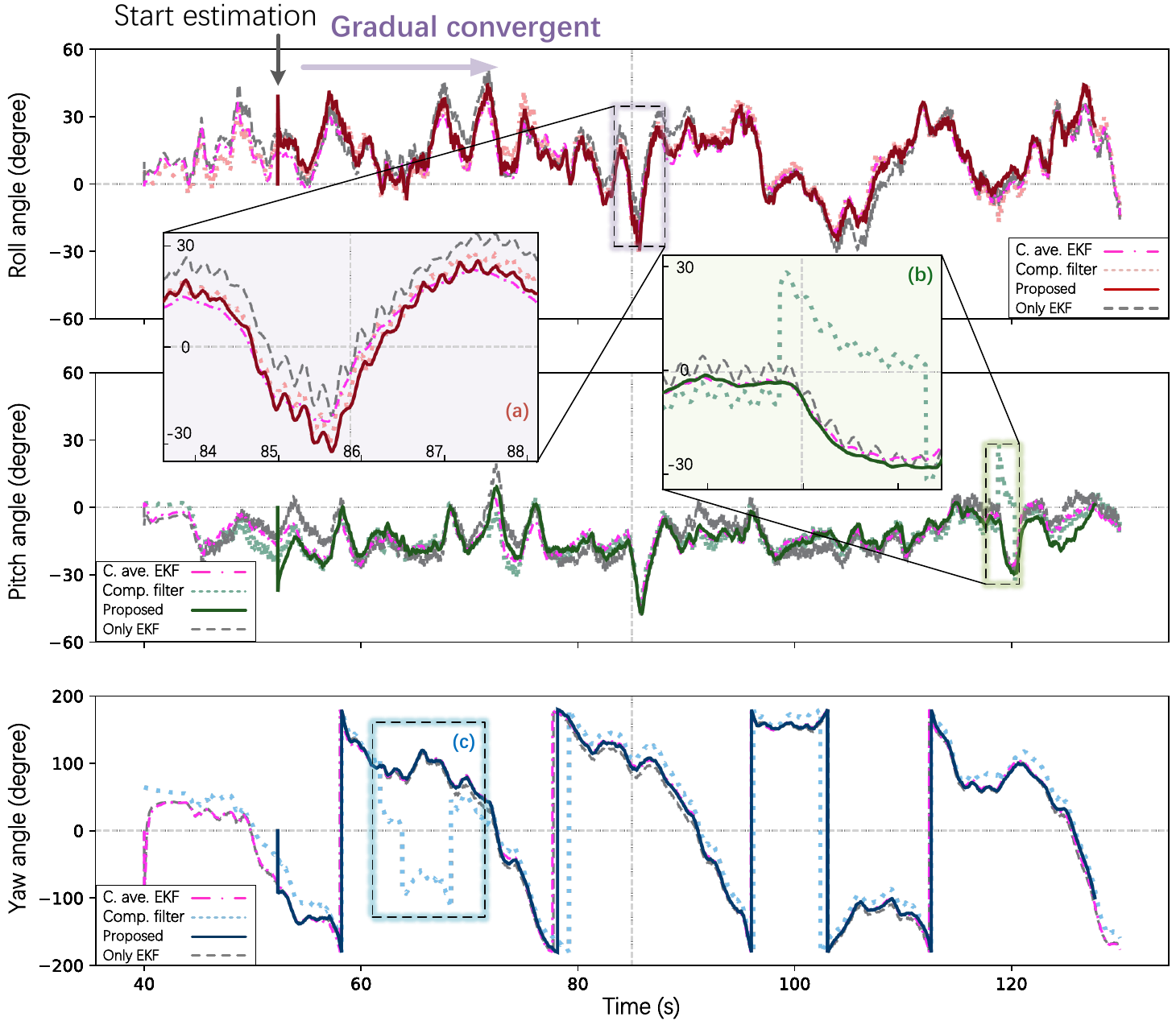}
  \caption{Attitude estimation algorithms comparison: In the legends, C. ave. EKF denotes the method using center-averaged data as input to the EKF, serving as an approximation of the ground truth. The comp. filter, used in PX4 for high-speed vehicles, is a lightweight variant of the Mahony complementary filter that recalibrates attitude estimates based on specific triggers, leading to the switching dynamics shown. Gray dashed lines represent the method employed in this work, where raw signals are straightforwardly input to the EKF.}
  \label{Fig:EKF_show}
\end{figure}
\begin{figure}[t]
  \centering
  \includegraphics[width=2.8in]{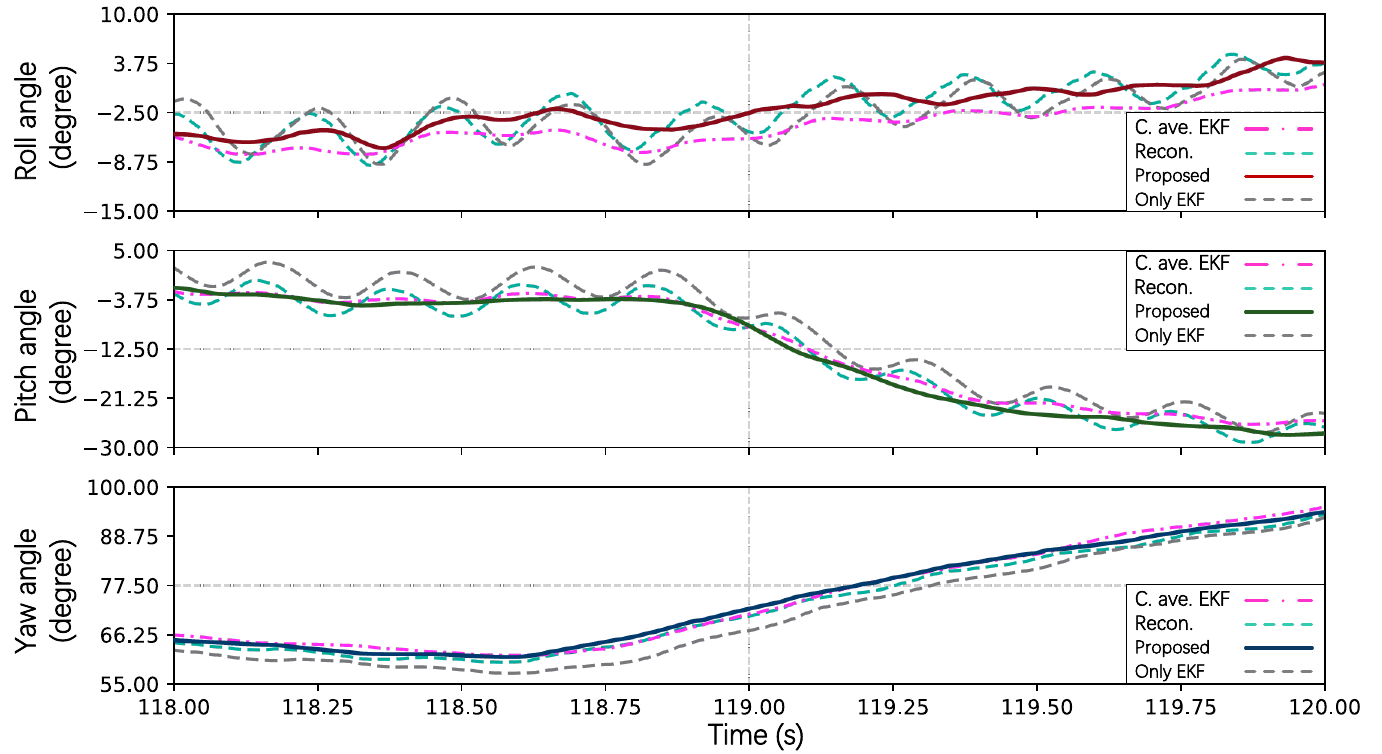}
  \caption{Reconstruction of the oscillating attitude from the oscillation-free signals and the online learned pattern:  The curves graphical representations are assigned similarly to those in Fig.~\ref{Fig:EKF_show}. 
  The reconstructed attitude curve is represented as teal dashed lines.}
  \label{Fig:EKF_showzoom}
\end{figure}
The oscillatory angular velocity patterns are also learned online, and their effects are significant. Notably, based on the curves illustrated in Fig.~\ref{Fig:Agv_res}-(a), the algorithm performs significantly better along the \(Y_B\)-axis than along the \(X_B\)-axis, primarily due to the greater susceptibility of rotations around \(X_B\) to disruptive airflow.
Subsequently, we conduct a comparative analysis to examine the rapid-rotation behavior of the FWAV under different wind conditions.
According to the learning performance shown in Fig.~\ref{Fig:Agv_res}-(b),
when a temporary side wind is exerted on the FWAV, 
the roll angular velocity drastically changes, during which the learning process is unable to adequately adapt.
Consequently, the oscillation-free data cease to exhibit relative smoothness in roll rotation.
It is noticeable that the motion corresponding to Fig.~\ref{Fig:Agv_res}-(c) also suffers a drastic roll rotation. 
Nonetheless, in this scenario, the absence of lateral wind allows the periodic pattern to maintain its current configuration, thereby allowing the oscillation-free data to maintain their inherent smoothness.
In conclusion, the algorithm leverages recent observations to build its oscillation recognition, allowing it to adapt to gradual changes while exhibiting sensitiveness to abrupt disturbances.
Although averaging over the last cycle can be computed online, 
it incurs a half-cycle delay compared to centered averaging.
Conversely, oscillation-free data from the proposed algorithm can be executed online and exhibit no obvious delay, as illustrated in Fig.~\ref{Fig:Agv_res}-(c). 
\subsection{Attitude Estimation}
The oscillation-free data obtained in last subsection are fed into the EKF attitude estimation algorithm discussed previously.
The corresponding results are explicitly shown in Fig.~\ref{Fig:EKF_show}.
The estimation process initiates when oscillation-free data becomes accessible, and the result from the developed method gradually convergent to the center average EKF results, which approximates the ground truth. 
In Fig.~\ref{Fig:EKF_show}-(a), the roll curve generated by the proposed method more closely aligns with the approximated ground truth, despite the presence of non-degraded oscillations.
Furthermore, the severe oscillations contribute to the significant estimation errors observed in both Fig.~\ref{Fig:EKF_show}-(b) and Fig.~\ref{Fig:EKF_show}-(c), which are effectively mitigated by the proposed method.
Another obvious characteristic of the proposed method is the elimination of the oscillatory component in the attitude estimation, as clearly demonstrated in Fig.~\ref{Fig:EKF_show}-(b). 
Since the periodic pattern is learned and recorded online, we can reconstruct the oscillating attitude from the oscillation-free estimation.
The obtained results are demonstrated in Fig.~\ref{Fig:EKF_showzoom}.
First, the reconstructed curve fluctuates around the oscillation-free estimation with an amplitude equivalent to that of the only-EKF estimation.
Second, the periodic oscillation generated by only-EKF estimation display a noticeable phase lag when compared to the reconstructed oscillating attitude.
These behaviors demonstrate that the proposed method effectively reconstructs the original oscillating attitude without delay, yielding a precise estimate of the current FWAV states, which can enhance future whole-body planning tasks.

\subsection{Internal-model-based State Estimation}
\begin{figure}[t]
  \centering
  \includegraphics[width=3.2in]{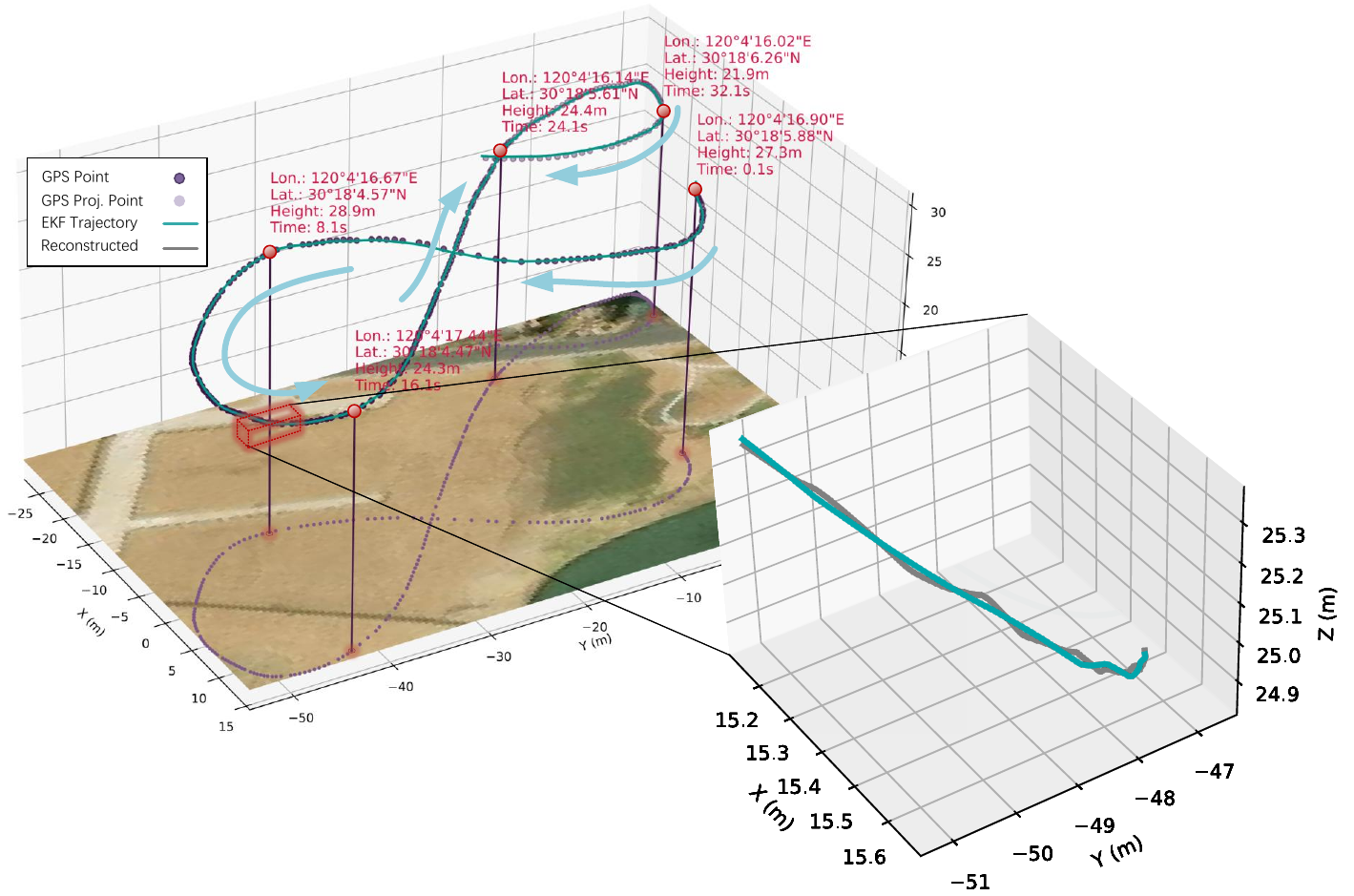}
  \caption{Internal-model-based state estimation: The GPS location data are highlighted in violet, with detailed information presented every 8 seconds. These data points are overlaid on a flat satellite map, depicted in lavender. Measurements along the $X$ and $Y$ axes are derived from the GPS coordinates using the ellipsoidal model.
 The estimated trajectory is represented in teal, and the reconstructed one in gray. 
  }
  \label{Fig:StateEKF}
\end{figure}

The internal-model-based EKF is developed to estimate slow-varying states, which consist of position, velocity and orientation,
utilizing oscillation-free measurements obtained previously.
In order to complete the internal model, we incorporate oscillation data for \(C_{L, \rm osc}\) and \(q_{\rm osc}\) as presented in equations \eqref{eq:C_T_inte_} and \eqref{eq:C_D_inte_}. 
The internal model outlined in equations \eqref{eq:internal_model} functions as predictive models incorporating information on both attitude and velocity, where the prediction of attitude is achieved from angular rate data.
Furthermore, GPS signals and attitude estimation play the observation roles during the update process.
To validate the effectiveness of our internal-model-based state estimation approach, we perform this state estimation utilizing a manual flight log. 
As illustrated in Fig.~\ref{Fig:StateEKF}, the estimated trajectory possesses higher resolution and aligns closely with that measured by RTK positioning. 
Using the obtained trajectory, we successfully reconstruct the oscillating original state by integrating the acceleration pattern.

\section{Conclusion}
A novel learning-based framework for state estimation in avian-inspired FWAVs is introduced in this paper. 
The method leverages low-cost MARG sensors and a divide-and-conquer strategy grounded in cycle-averaged aerodynamics to separate high-frequency oscillatory disturbances from slow-varying dynamics. Experimental evaluations demonstrate that online periodic pattern learning—implemented via FFT, cross-correlation, and real-time clustering—yields oscillation-free signals that considerably enhance attitude estimation accuracy. 
When integrated into an internal-model-based EKF, the refined measurements yield state estimates that closely correspond to high-precision positioning data, thereby improving overall flight stability. 
This work contributes to the agile operation of autonomous FWAVs in complex environments, with further exploration to be addressed in our forthcoming studies.
\bibliographystyle{IEEEtran}
\bibliography{root}

\begin{IEEEbiography}[{\includegraphics[width=1in,height=1.25in,clip,keepaspectratio]{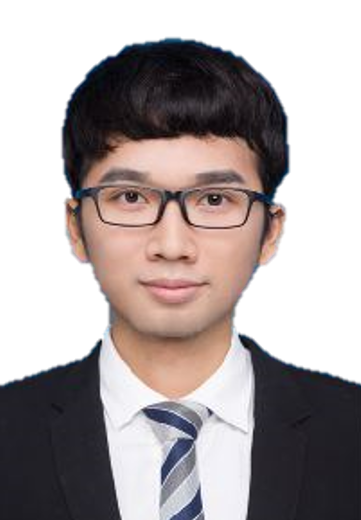}}]
{Chen Qian} received the B.S. degree in intelligence science and technology from Nankai University, Tianjin, China, in 2015, and the Ph.D. degree from the College of Artificial Intelligence, Nankai University, in 2021.

He is currently an assistant researcher with the Center for X-mechanics of Zhejiang University. His research interests include the design and control of flapping wing vehicles and other bionic robots.
\end{IEEEbiography}

\begin{IEEEbiography}[{\includegraphics[width=1in,height=1.25in,clip,keepaspectratio]{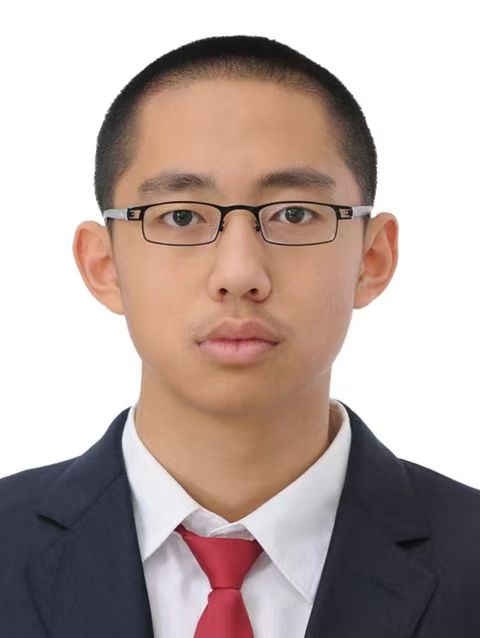}}]
{Jiaxi Xing} received the B.S. degree in aeronautical science and technology in 2020 from Beihang University, Beijing, China.

He is currently working at Huzhou Institute of Zhejiang University. His research interest includes autonomous navigation of flapping-wing robots.
\end{IEEEbiography}

\begin{IEEEbiography}[{\includegraphics[width=1in,height=1.25in,clip,keepaspectratio]{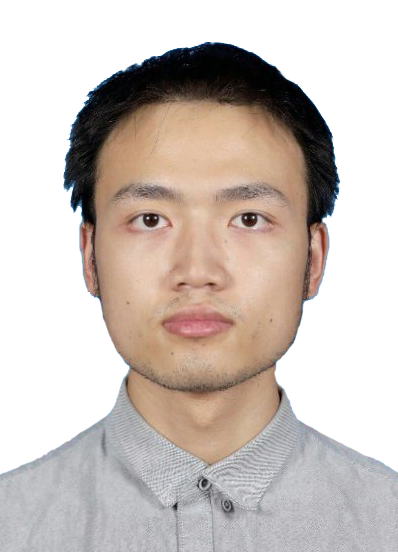}}]
{Jifu Yan} received the B.S. degree in intelligent science and technology in 2023 from Nankai University, Tianjin, China.

He is currently working toward the Master degree in artificial intelligence with the Institute of Robotics and Automatic Information System. His research interest includes nonlinear control of flapping-wing robots.

\end{IEEEbiography}

\begin{IEEEbiography}[{\includegraphics[width=1in,height=1.25in,clip,keepaspectratio]{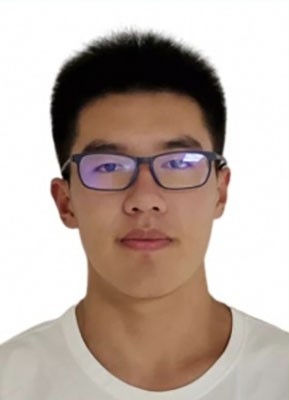}}]
{Mingyu Luo} is currently pursuing the B.S. degree in mechanical engineering from Zhejiang University, Hangzhou, China. 

He received funding from the National Student Research Training Program to conduct research on the structural optimization of biomimetic flapping-wing robots. His research interests include flapping-wing robots, mechanical design and manufacture.
\end{IEEEbiography}

\begin{IEEEbiography}[{\includegraphics[width=1in,height=1.25in,clip,keepaspectratio]{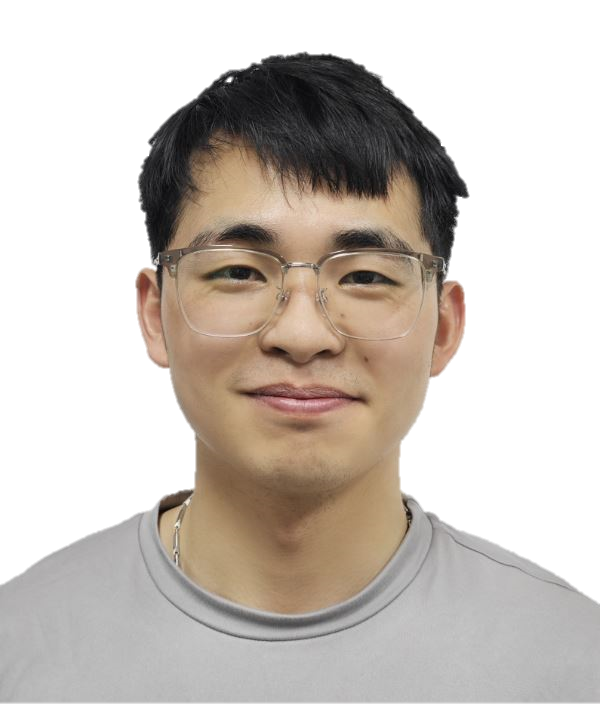}}]
{Shiyu Song} received the M.S degree in School of Aeronautics in 2024 from Northwestern Polytechnical University, Xi'an, China.

He is currently pursuing the Ph.D degree in mechanical engineering from Zhejiang University, Hangzhou, China. His research interests include flapping-wing robot, quadcopter control, computer vision.
\end{IEEEbiography}

\begin{IEEEbiography}[{\includegraphics[width=1in,height=1.25in,clip,keepaspectratio]{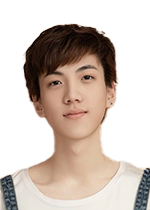}}]
{Xuyi Lian} is currently pursuing the B.S. degree in aircraft design and engineering with a specialization in aircraft information and electronics from Zhejiang University, Hangzhou, China. 

He received funding from the National Student Research Training Program to conduct research on the structural optimization of biomimetic flapping-wing robots. His research interests include flapping-wing robots, soft robotics, and control systems.

\end{IEEEbiography}

\begin{IEEEbiography}[{\includegraphics[width=1in,height=1.25in,clip,keepaspectratio]{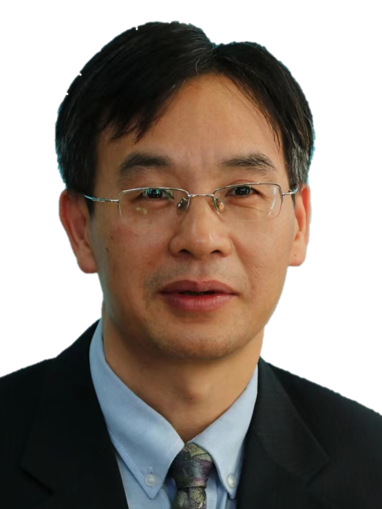}}]
{Yongchun Fang} received the B.S. and M.S. degrees in control theory and applications from Zhejiang University, Hangzhou, China, in 1996 and 1999, respectively, and the Ph.D. degree in electrical engineering from Clemson University, Clemson, SC, USA, in 2002.,From 2002 to 2003, he was a Postdoctoral Fellow with the Sibley School of Mechanical and Aerospace Engineering, Cornell University, Ithaca, NY, USA.

He is currently a Full Professor with the Institute of Robotics and Automatic Information Systems, College of Artificial Intelligence, Nankai University, Tianjin, China. His research interests include nonlinear control, visual servoing, control of underactuated systems, and atomic-force-microscopy-based nanosystems.
\end{IEEEbiography}

\begin{IEEEbiography}[{\includegraphics[width=1in,height=1.25in,clip,keepaspectratio]{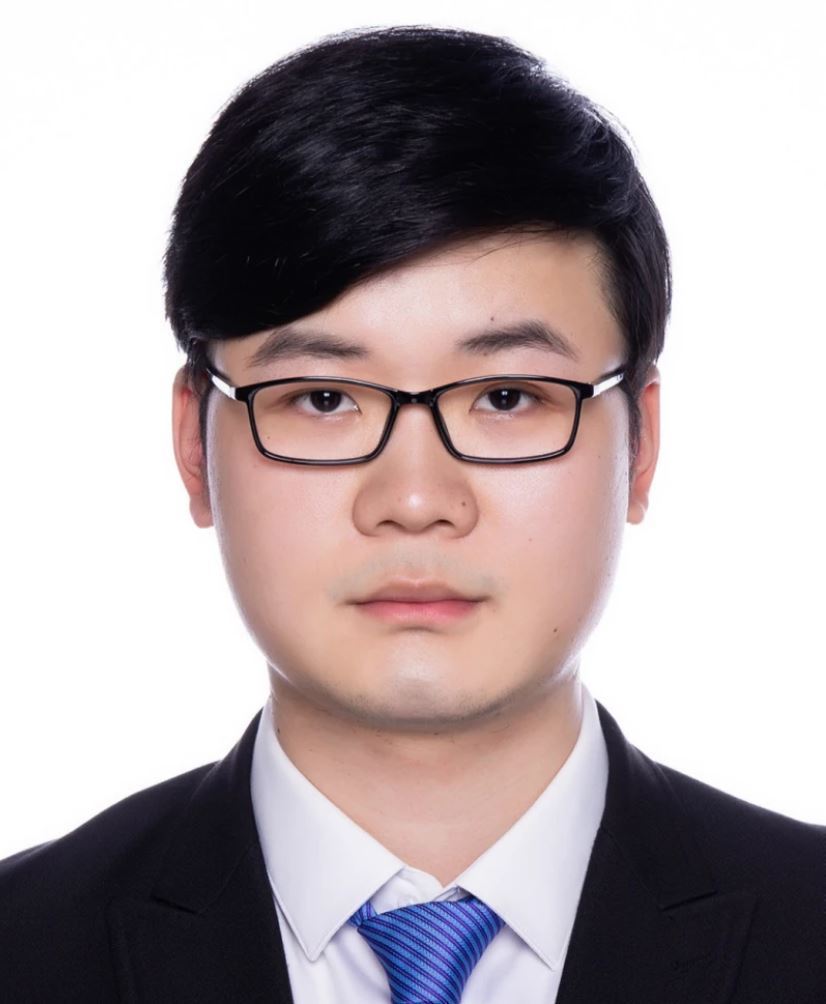}}]
{Fei Gao} received the Ph.D. degree in electronic and computer engineering from the Hong Kong University of Science and Technology, Hong Kong, in 2019.

He is currently a tenured associate professor at
the Department of Control Science and Engineering, Zhejiang University, where he leads the Flying Autonomous Robotics (FAR) group affiliated
with the Field Autonomous System and Computing
(FAST) Laboratory. His research interests include
aerial robots, autonomous navigation, motion planning, optimization, and localization and mapping.
\end{IEEEbiography}

\begin{IEEEbiography}[{\includegraphics[width=1in,height=1.25in,clip,keepaspectratio]{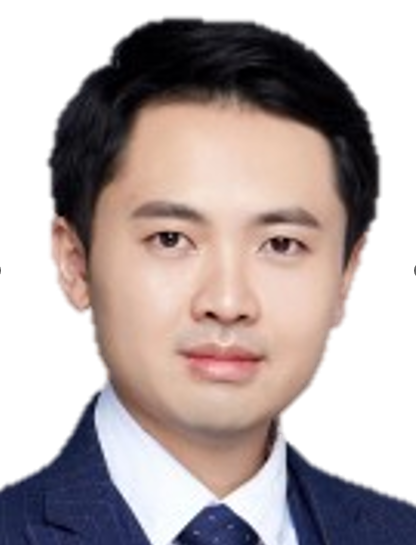}}]
{Tiefeng Li}  received the B.E. and Ph.D. degrees from Zhejiang University, Hangzhou, China, in 2007 and 2012, respectively.

He is currently a Professor with School of Aeronautics and Astronautics of Zhejiang University. His research interests include smart materials, soft robotics and bionic robots.
\end{IEEEbiography}

\end{document}